# Exact Solution for the Protected TEM edge mode in a PTD-Symmetric Parallel-Plate Waveguide

E. Martini, *Senior Member*, *IEEE*, M. G. Silveirinha *Fellow, IEEE*, S. Maci, *Fellow, IEEE*

*Abstract*— A Parity Time-reversal Dual (PTD) symmetric structure constituted by a Perfectly-Electric-Perfectly magnetic (PEC-PMC) parallel plate waveguide (PPW) is analyzed. This waveguide supports unimodal transverse electromagnetic (TEM) edge mode propagation protected against back-scattering from a certain class of deformations and defects. The TEM solution is found in analytical form by using three different methods, namely conformal mapping, mode-matching, and Fourier-transform methods. It is shown through numerical simulations that the mode propagation is robust with respect to deformations such as 90° bends and discontinuity such as transition to free-space. Implementation of the PMC boundary conditions via both a bed of nails and a mushroom structure is also successfully investigated.

*Index Terms*— Metasurface, Topological modes, Parity-time-duality symmetry

## I. INTRODUCTION

Topological Edge Modes (TPEMs) are the electromagnetic (EM) counterpart of the edge states in the integer quantum Hall effect that occurs due to topological phase transitions of matter, the discovery at the origin of the Nobel Prize awarded to Thouless, Haldane and Kosterlitz in 2016 [1]. Topological methods, originally developed for electronic systems, have been generalized to electromagnetic systems, opening new and unexpected opportunities for innovation [2]-[4]. The novel aspect of TPEMs is that they can be unidirectional and thereby protected against back-scattering; thus, they may enable a wave-guiding immune to the undesired effects of reflections due to disorder, imperfections, obstacles or deformations of the propagation path. TPEM protection normally requires non-reciprocal elements [2]-[7]; however, recent researches have shown it can be also generated by reciprocal materials and thereby topological protected edge modes may arise in time-reversal invariant structures [8]-[13]. In this case, the edge modes are bidirectional, but largely (in ideal cases *totally*) immune from backscattering due to chiral-type properties.

Furthermore, a few recent works have highlighted that a broad class of (non-topological) reciprocal systems may offer as well protection against back-scattering [14]-[17]. In this case, the key property that guarantees that some propagating mode is immune to reflections is a duality link between the constitutive parameters of the relevant materials [14]-[17]. In particular, it was first shown in [16] that a wide family of bi-directional (not necessarily reciprocal) *N*-port networks invariant under the combined action of the parity (P), time-reversal (reciprocity) (T) and duality (D) operators, are characterized by a scattering matrix with $s_{11} = .... = s_{NN} = 0$. Thus, a PTD-invariant microwave network is always matched at all ports. Thereby, PTD-invariant platforms may support waves that are insensitive to any form of perturbations or defects that do not break the PTD symmetry. This scattering anomaly (i.e., the absence of back-scattering in a bi-directional system) is the optics analogue of the quantum-spin-Hall effect in quantum mechanics [16]. Note that the number of ports is not necessarily identical to the number of physical waveguides because it may happen that a certain guide supports multiple modes. The condition $s_{ii} = 0$, guarantees that the energy coupled to a generic *single-mode* waveguide is rerouted to the other waveguides with no back-reflections. When a waveguide supports two or more propagating modes some of the energy coupled to it may return-back due to modal conversion. As detailed in [16], when a waveguide supports an *odd* number, e.g., 1, 3, 5, ..., of propagating modes it is always possible to find some suitable excitation that guarantees that no power is returned-back via modal conversion [16].

Different from topological systems, the PTD invariance is not a global property but rather a single-frequency condition. Rather remarkably PTD-systems can be formed by reciprocal materials [15]-[16]. Hence, the PTD symmetry only requires reciprocal metamaterials, and this renders the fabrication quite convenient. The relative permittivity and permeability of a PTD-invariant system are linked to the constitutive permittivity ($\bar{\varepsilon}$) and permeability ($\bar{\mu}$) tensors as $\bar{\varepsilon}(x,y,z) = \mathbf{V} \cdot \bar{\mu}^T(x,y,-z) \cdot \mathbf{V}$, where the superscript T denotes transpose and $\mathbf{V}$ is a tensor represented by a diagonal matrix with diagonal elements $\{1,1,-1\}$ [16]. Furthermore, the magneto-electric tensor $\bar{\xi}$, associated with a bianisotropic response, if not zero, must satisfy the condition $\bar{\xi}(x,y,z) = -\mathbf{V} \cdot \bar{\xi}^T(x,y,-z) \cdot \mathbf{V}$ [16]. In the previous formulas, it is implicit that the parity transformation is $(x,y,z) \to (x,y,-z)$ but other choices are possible. Note that the topological insulators introduced by Khanikaev *et al* [10] are particular examples of PTD-symmetric systems with an Ω-type bianisotropic coupling [16]. Furthermore, the edge waveguide described in [18] is also PTD-invariant.

In this paper, we analyze in detail the fundamental mode of a (non-bianisotropic) PTD-symmetric guide formed by pairing two dual parallel-plate waveguides with PEC and PMC walls. This structure was first proposed and numerically investigated in [15]. Here we obtain an analytical exact solution for the fundamental mode. While the two individual waveguides exhibit cut-off bandwidths from zero frequency to the frequency at which the distance of the walls is a quarter of wavelength, their pairing generates a transverse electromagnetic (TEM) mode protected with respect to PTD-type defects. The resulting PTD-TEM mode is strongly confined along the discontinuity of the boundary conditions with exponential penetration of the order of the distance between the walls.

The paper is organized as follows. In Section II, three forms of the exact solution for this protected TEM mode are derived, based on *i)* a conformal transformation, *ii)* mode-matching and *iii)* a Fourier-transform method. In section III, examples of segmented protected propagation and radiation through an open-ended termination are presented. In Section IV a design of the PMC walls using bed of nails and mushroom structure is studied by a full-wave analysis to show possible practical implementations. Conclusions are drawn in Section V.

## II. CANONICAL SOLUTION

The transverse cross-section of the PTD parallel-plate waveguide (PTD-PPW) is shown in Fig. 1a. In this article, we adopt $(x, y', z) \to (x, -y', z)$ as the relevant parity transformation so that the PTD condition reduces to $\varepsilon(x, y', z) = \mu(x, -y', z)$ for structures formed by simple non-bianisotropic reciprocal isotropic materials. Here, $y' = y - y_0$ with $y_0$ being the coordinate of the symmetry plane. Each waveguide wall is formed by PMC-PEC boundaries, and the center of symmetry is taken as the middle plane $y_0 = d/2$, being $d$ the height of the waveguide. Each individual PPW exhibits a bandgap from zero frequency to the cut-off frequency for which $d = \lambda/4$.

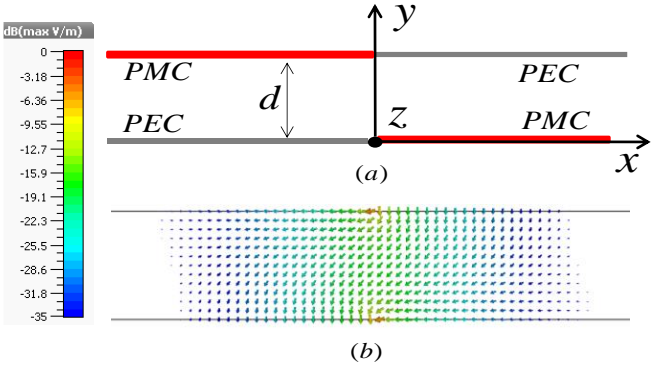

Fig. 1 (a) Geometry of the canonical problem; (b) distribution of the transverse electric field obtained from CST Microwave Studio for the dominant TEM mode.

To figure out the EM-field structure we are going to find, Figure 1b shows the distribution of electric field lines obtained by CST for the dominant TEM mode. Let us introduce the static electric potential $\psi(x, y)$, the gradient of which represents the electric field in the transverse plane: $\nabla_t \psi(x, y) = -\mathbf{e}_t(x, y)$ with $\nabla_t^2 \psi(x, y) = 0$. The boundary conditions impose that

$$\frac{\partial}{\partial y}\psi(x,0) = 0 \quad \text{for } y=0, \quad x>0 \quad (1)$$

$$\frac{\partial}{\partial y}\psi(x,d) = 0 \quad \text{for } y=d, \quad x<0 \quad (2)$$

$$\psi(x,0) = -V_0/2 \quad \text{for } y=0, \quad x<0 \quad (3)$$

$$\psi(x,d) = V_0/2 \quad \text{for } y=d, \quad x>0 \quad (4)$$

where $V_0$ is the difference of potential between the top and bottom PEC plates. The last two equations impose the potential $\pm V_0/2$ on the two PEC parts of the waveguide.

### A. Conformal Mapping

The electrostatic problem can be solved by transforming the original domain into a domain in which the solution is known in a simple form. Using a combination of two Schwarz-Christoffel transformations it is indeed possible to map the original problem in Fig. 1 (in which for simplicity, $d=1$) into a tilted square with lateral side equal to 1 (Fig. 2).

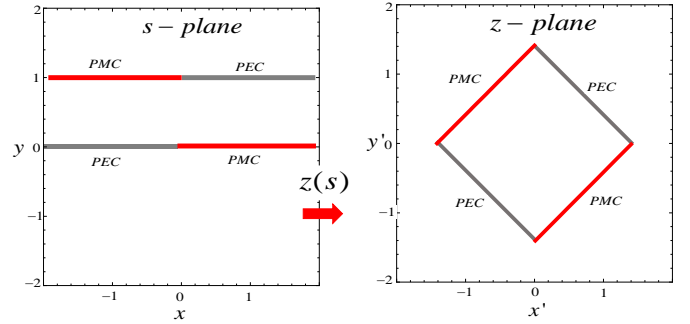

Fig. 2 Transformation of the domain by the function z(s)

In this domain, the potential is simply obtained by

$$\phi(z) = V_0 \operatorname{Re}\left\{z e^{-j\pi/4}\right\} \quad (5)$$

where $z = x' + jy'$ are the coordinates of the transformed domain. The relevant conformal mapping from the original domain $0 < \operatorname{Im}\{s\} < d = 1$ into the square is given by:

$$z(s) = e^{j\pi/4}\left[\frac{jF_2(\xi(s))}{F_{-1}\left(\frac{\pi}{2}\right)} - (1+j)\frac{1}{2}\right], \quad (6)$$

where $\xi(s) = \arcsin\left(\left(1 - \coth\left(\frac{\pi s}{2}\right)\right)^{-1/2}\right)$ and

$$F_m(\xi) = \int_0^\xi d\theta \frac{1}{\sqrt{1 - m\sin^2\theta}} \quad (7)$$

is the incomplete elliptic integral of the first kind. Finally, one can find the solution as $\psi(x,y) = \phi(z(s/d))$ with $s=x+jy$. This leads to the following form of the potential

$$\psi(x,y) = V_0 \operatorname{Re}\left[\frac{j}{F_{-1}\left(\frac{\pi}{2}\right)} \int_0^{\xi\left(\frac{x}{d}+j\frac{y}{d}\right)} \frac{1}{\sqrt{1-2\sin^2\theta}} d\theta\right] - \frac{V_0}{2} \quad (8)$$

$$\xi(x+jy) = \arcsin\left(\left(1 - \coth\left(\frac{\pi(x+jy)}{2}\right)\right)^{-1/2}\right) \quad (9)$$

Figure 3 shows the equipotential contour lines for the case $d=1$ (note that the behavior for a generic value of $d$ is obtained by normalizing both the space coordinates by $d$).

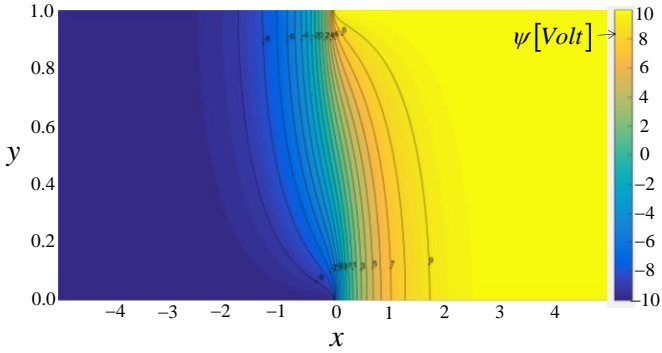

Fig. 3 Distribution of the potential $\psi(x,y)$ for $d=1$ and $V_0=20$ [V] calculated through the conformal mapping solution [Eq. (8)].

### B. Mode matching

The two-part problem can be solved by mode-matching at $x=0$ after expanding the potential in both regions, $x<0$ and $x>0$, in terms of a complete set of functions which respect the boundary conditions and the Laplace equation.

For the sake of convenience, we construct the solution in terms of the discontinuous potential $\varphi(x,y)$ defined as

$$\varphi(x,y) = \psi(x,y) - \text{sgn}(x) V_0 / 2. \quad (10)$$

where $\text{sgn}(x)$ is the function which is 1 for $x$ positive and $-1$ for $x$ negative. Although the introduction of the discontinuous potential $\varphi(x,y)$ is not essential in finding the solution, it simplifies the following derivation because it satisfies conventional Dirichlet and Neumann boundary conditions. Indeed, according to (1)-(4), the function $\varphi(x,y)$ respects

$$\frac{\partial}{\partial y}\varphi(x,0) = 0 \quad \text{for } y=0, \quad x>0 \quad (11)$$

$$\frac{\partial}{\partial y}\varphi(x,d) = 0 \quad \text{for } y=d, \quad x<0 \quad (12)$$

$$\varphi(x,0) = 0 \quad \text{for } y=0, \quad x<0 \quad (13)$$

$$\varphi(x,d) = 0 \quad \text{for } y=d, \quad x>0 \quad (14)$$

Furthermore, $\varphi$ goes to zero for $|x| \to \infty$. Also, $\varphi(x,y)$ respects $\nabla^2 \varphi(x,y) = -\nabla^2 [\text{sgn}(x) V_0 / 2]$, that is

$$\nabla_t^2 \varphi(x,y) = -V_0 \delta'(x) \quad (15)$$

for $x \neq 0$, $\nabla_t^2 \varphi(x,y) = 0$. Hence, for each of the relevant domains ($x<0$ or $x>0$) $\varphi(x,y)$ can be expanded into a set of functions satisfying $\nabla_t^2 \varphi_n(x,y) = 0$ and the relevant boundary conditions. A complete set of functions with these properties is

$$\begin{aligned} \varphi_n^s(x,y) &= \sin(\alpha_n y) e^{-\alpha_n |x|} & x<0 \\ \varphi_n^c(x,y) &= -\sin(\alpha_n (d-y)) e^{-\alpha_n |x|} & x>0 \end{aligned} \quad n=0,1,2\ldots \quad (16)$$

with $\alpha_n d = (n\pi + \pi/2)$. Note that we can also write $\varphi_n^c(x,y) = (-1)^{n+1}\cos(\alpha_n y)e^{-\alpha_n |x|}$. The function $\varphi(x,y)$ is therefore represented as

$$\varphi(x,y) = \sum_{n=0}^{\infty} a_n \varphi_n^s(x,y) \qquad x<0 \quad (17)$$

$$\varphi(x,y) = \sum_{n=0}^{\infty} b_n \varphi_n^c(x,y) \qquad x>0 \quad (18)$$

From the symmetry of the problem, it is necessary that $\varphi(x,y) = -\varphi(-x, d-y)$. Hence, observing that $\varphi_n^s(x,y) = -\varphi_n^c(-x, d-y)$, we find that $b_n = a_n$. Since the potential $\psi(x,y)$ in (10) is continuous, $\varphi(x,y)$ is discontinuous at $x=0$, and its discontinuity is equal to $V_0$. This implies the condition

$$\sum_{n=0}^{\infty} a_n \{\sin(\alpha_n y) + \sin[\alpha_n(d-y)]\} = V_0. \quad (19)$$

Similar conditions are found by imposing the continuity of the derivative with respect to $x$ and $y$:

$$\sum_{n=0}^{\infty} a_n \alpha_n \{\sin(\alpha_n y) - \sin[\alpha_n(d-y)]\} = 0, \quad (20)$$

$$\sum_{n=0}^{\infty} a_n \alpha_n \{\cos(\alpha_n y) - \cos[\alpha_n(d-y)]\} = 0. \quad (21)$$

Projecting both the right- and left-hand sides of (19), (20) and (21) onto the functions $\sin(\alpha_m y)$ yields three infinite linear systems ($i=0,1,2$)

$$\sum_{n=0}^{\infty} \zeta_{mn}^{(i)} a_n = V_m^{(i)}, \quad m=0,1,2\ldots \quad (22)$$

where the expressions of $\zeta_{mn}^{(i)}$ and $V_m^{(i)}$ are given in the Appendix (sec. A). On the basis of the analytical form expressions of these coefficients, given in (44)-(48) of Appendix A, and after some algebraic manipulations of (22) which do not invoke other properties of the boundary value problem, one obtains

$$\sum_{n=0}^{\infty} a_{2n} \frac{4n+1}{2n-(2m-1)} = 0 \quad m=1,2,3\ldots \quad (23)$$

$$\sum_{n=0}^{\infty} a_{2n} = V_0 \quad (24)$$

$$a_{2n+1} = 0 \qquad n=0,1,2\ldots \quad (25)$$

From (23)-(25) the coefficients $a_n$ can be found in a closed form following the residue-calculus method suggested in [19, p. 647 ff.]. According to this method, one needs to find a meromorphic function $f(z)$ of the complex variable $z$, with zero-residue at infinity, such that (23) can be interpreted as the vanishing summation of the residues of $f(z)/[\exp(jz\pi)+1]$. The function $f(z)$ can be written as $f(z) = \sum_{n=0}^{\infty} a_{2n} \frac{4n+1}{z-2n}$; it has poles at $z=2n$ with residues $a_{2n}(4n+1)$, for $n=0,1,2,\ldots$ and zeros at $z = (2m-1)$, for $m=1,2,3,...$ to cancel the poles of $1/(\exp(jz\pi)+1)$ and satisfy (23); furthermore, it should go to zero at infinity. In Appendix (sec. B) it is shown that $f(z)$ have the form

$$f(z) = C \frac{\Gamma\left(-\frac{z}{2}\right)}{\Gamma\left(-\frac{z}{2}+\frac{1}{2}\right)} \quad (26)$$

where $\Gamma(s)$ is the Gamma function and $C$ is an arbitrary constant eventually determined through (24). The residues of this function are given by (see Appendix, sec. B)

$$R_n = KV_0 \frac{(2n)!}{(n!)^2 4^n} \quad n=0,1,2,... \quad (27)$$

where $K$ is related to $C$ by $KV_0 = 2C/\sqrt{\pi}$. The coefficients $a_{2n}$ can be found through (22) equating $a_{2n}(4n+1) = R_n$, i.e.,

$$a_{2n} = KV_0 \frac{(2n)!}{(4n+1)(n!)^2 4^n} \quad n=0,1,2,... \quad (28)$$

where the constant $K$ can be found imposing (24), namely

$$K = \left[\sum_{n=0}^{\infty} \frac{(2n)!}{(4n+1)(n!)^2 4^n}\right]^{-1} \approx 0.7965 \quad (29)$$

The above coefficients, inserted into (17)-(18) give the electric potential in closed analytical form. With the substitution $n \to 2n$ one gets the final solution as

$$\begin{aligned}\varphi(x,y) &= V_0 \sum_{n=0}^{\infty} c_n e^{-\xi_n|x|}\left[\sin(\xi_n y)\,u(-x) - \cos(\xi_n y)u(x)\right] \\ \psi(x,y) &= \varphi(x,y) + \mathrm{sgn}(x)V_0/2 \\ \xi_n d &\equiv \alpha_{2n}d = 2\pi n + \frac{\pi}{2} \\ c_n &\equiv \frac{a_{2n}}{V_0} = 0.7965\frac{(2n)!}{(4n+1)(n!)^2 4^n} \quad n=0,1,2,...\end{aligned} \quad (30)$$

where $u(x)$ is Heaviside unit step function, which is equal to 1 for $x$ positive and vanishes for $x$ negative. Note that $\sum_{n=0}^{\infty} c_n = 1$.

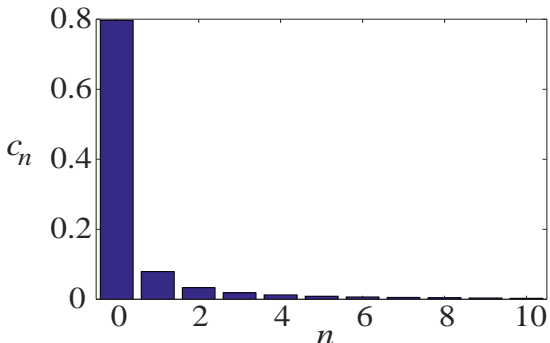

Fig. 4 Distribution of the first 10 coefficients $c_n$ in Eq. (30).

The distribution of the coefficients $c_n$ is shown in Fig. 4. It can be seen that the series is strongly dominated by the first coefficient, that is by the modal function with $n=0$ (the first coefficient is 80% of the entire summation, which is unitary).

As a check of the result, the continuous potential $\psi(x,y)$ has been numerically calculated from (30) and compared with the one provided by (8). The results are shown in Fig. 5. As expected, the two representations provide the same result.

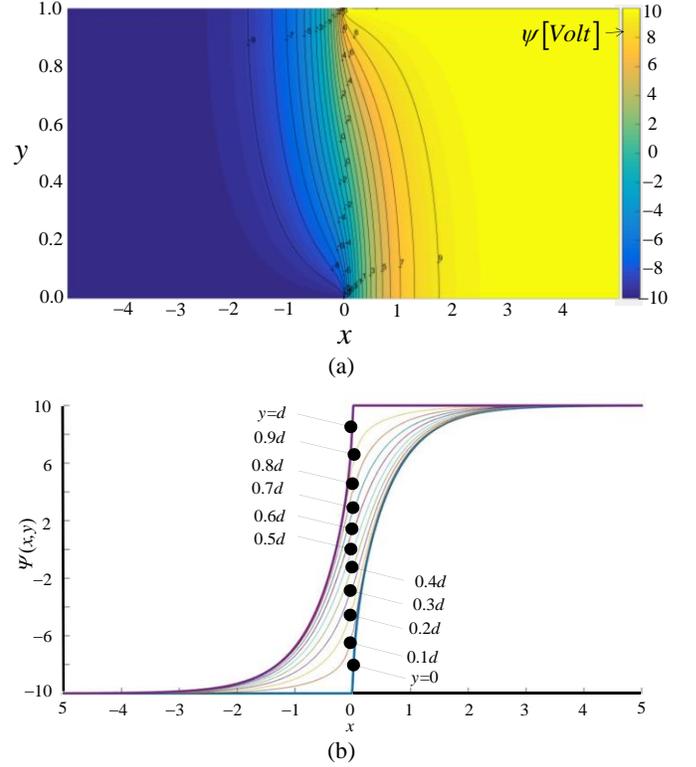

Fig. 5 (a) Distribution of the potential $\psi(x,y)$ for $d=1$ and $V_0=20$ [V] calculated through (30). This distribution is identical to the one obtained in Fig. 3 from the conformal mapping. (b) distribution of the potential $\psi(x,y)$ as a function of $x$ for various values of y.

### C. Solution in the Fourier-domain

We can solve the problem in the Fourier-transform domain, by introducing

$$\Phi(k_x,y) = \int_{-\infty}^{\infty} \varphi(x,y)e^{jk_x x}dx \quad (31)$$

Due to (15), $\Phi(k_x,y)$ satisfies

$$-k_x^2 \Phi(k_x,y) + \frac{\partial^2 \Phi}{\partial y^2}(k_x,y) = V_0 j k_x \quad (32)$$

The function $\varphi(x,y)$ is absolutely integrable in $x$ and hence $\Phi(k_x,y)$ is finite for $k_x = 0$. Furthermore, $\varphi(x,y)$ is discontinuous at the origin, and hence its Fourier transform should decay as $1/k_x$.

A suitable form for the spectrum in (31) can be derived by invoking the symmetry of the problem; this spectrum is subjected to the symmetry relation $\Phi(k_x,y) = -\Phi(-k_x, d-y)$. A solution of (32) that respects this symmetry is

$$\Phi(k_x,y) = \frac{-V_0}{jk_x}\left[A(k_xd)\frac{\sinh(k_xy)}{\sinh(k_xd)} + A(-k_xd)\frac{\sinh[k_x(d-y)]}{\sinh(k_xd)} - 1\right] \quad (33)$$

with $A(0) = 1$. It is apparent that

i) under condition $A(0) = 1$ the pole in $k_x=0$ is cancelled for any $y$ in agreement with the fact that $\varphi(x,y)$ is integrable in $x$;

ii) since the function $\varphi(x,y)$ is discontinuous at the origin as $\text{sgn}(x)V_0/2$, $\Phi(k_x,y)$ for $k_x \to \infty$ tends to $V_0/(jk_x)$, namely $\lim_{k_x \to \infty} jk_x\Phi(k_x,y) = V_0$ for any $y$. This implies that the sum of the first two terms inside the square brackets, and therefore $A(k_xd)$, should approach zero for $k_x \to \infty$.

iii) the spectrum $\Psi(k_x,y)$ of the continuous function $\psi(x,y)$ can be obtained just neglecting the last unity term inside the square brackets in (33).

In order to find $A(s)$, we assume that $\varphi(x,y)$ has an expansion as in the first eq. in (30), without any a priory assumption on the expression of the coefficients $c_n$. When this expansion is specialized for $y=0$ and $y=d$ in (31), one obtains

$$\Phi(k_x,0) = -V_0\sum_{n=0}^{\infty} c_n \int_0^{\infty} e^{-\xi_n x}e^{jk_x x}dx = \sum_{n=0}^{\infty}\frac{V_0 c_n}{-\xi_n + jk_x}$$

$$\Phi(k_x,d) = V_0\sum_{n=0}^{\infty} c_n \int_{-\infty}^0 e^{\xi_n x}e^{jk_x x}dx = \sum_{n=0}^{\infty}\frac{V_0 c_n}{\xi_n + jk_x} \quad (34)$$

The poles of $\Phi(k_x,0)$ are all in the lower half-space (LHP) of the complex $k_x$ plane since the function should be analytic in the UHP (reversely for $\Phi(k_x,d)$). The Fourier integral for a generic $y$ should be calculated as

$$\Phi(k_x,y) = V_0\sum_{n=0}^{\infty} c_n\left[\frac{\sin(\xi_n y)}{\xi_n + jk_x} + \frac{\cos(\xi_n y)}{-\xi_n + jk_x}\right] \quad (35)$$

which is a form that automatically respects (32), under (19) and (20). One can easily see that the function $A(\pm k_xd)$ in (33) is related to $\Phi(k_x,0)$ and $\Phi(k_x,d)$ by the equation $-V_0A(\pm k_xd)/(jk_x) = \Phi(k_x,d;0) - V_0/(jk_x)$. Therefore, using (34), and observing that $\lim_{k_x \to \infty} jk_x\Phi(k_x,d) = V_0$ (see point ii above) implies $\sum_{n=0}^{\infty} c_n = 1$, one has

$$A(k_xd) = -jk_x\sum_{n=0}^{\infty}\left[\frac{c_n}{\xi_n + jk_x} - \frac{1}{jk_x}\right] = \sum_{n=0}^{\infty}\frac{c_n\xi_n}{(\xi_n + jk_x)} \quad (36)$$

We stress that the derivation of the last term in (36) does not require a particular form of the coefficients $c_n$, thus rendering this third method of solution independent of the second method described in Section II B. By comparing the last term in (36) with (23) one can argue that

$$\begin{array}{l}A(s) \text{ has } \textit{zeros} \text{ at } s = z_m = j(2\pi m - \frac{\pi}{2}) \quad m=1,2,3...\\ A(s) \text{ has } \textit{simple poles} \text{ at } s=j\xi_nd = j(2\pi n + \frac{\pi}{2}) \quad n=0,1....\end{array} \quad (37)$$

From these properties, one can find $A(s)$ in terms of the following combination of Gamma functions (see Appendix, sec. C)

$$A(s) = \sum_{n=0}^{\infty}\frac{c_n\xi_nd}{(\xi_nd + js)} = \frac{\Gamma(3/4)}{\Gamma(1/4)}\frac{\Gamma\left(j\frac{s}{2\pi} + \frac{1}{4}\right)}{\Gamma\left(j\frac{s}{2\pi} + \frac{3}{4}\right)} \quad (38)$$

where the leading normalization coefficient is found imposing that $A(0) = 1$ ($\Gamma(1/4) = 3,6256$, $\Gamma(3/4) = 1,2254$). The Gamma function in the numerator constructs the poles of (38) and the one at the denominators constructs the zeros. Through the general expression $\Gamma(z+a)/\Gamma(z+b) \sim z^{a-b}$, the behavior at infinity can be found to be $s^{-1/2}$, which goes to zero, as expected. Upon insertion of (35) in the inverse FT,

$$\varphi(x,y) = \frac{1}{2\pi}\int_{-\infty}^{\infty}\Phi(k_x,y)e^{-jk_xx}dk_x \quad (39)$$

one can recover (30) through the Jordan Lemma applied to (39), just summing up the residues in the LHP for $x$ positive and in UHP for $x$ negative. The substitution of (38) in (33) allows for having a closed form expression for the spectrum $\Phi(k_x,y)$.

Figure 6 shows the behavior of the spectrum of the potential for some values of $y$. The curves obtained by the different methods are coincident.

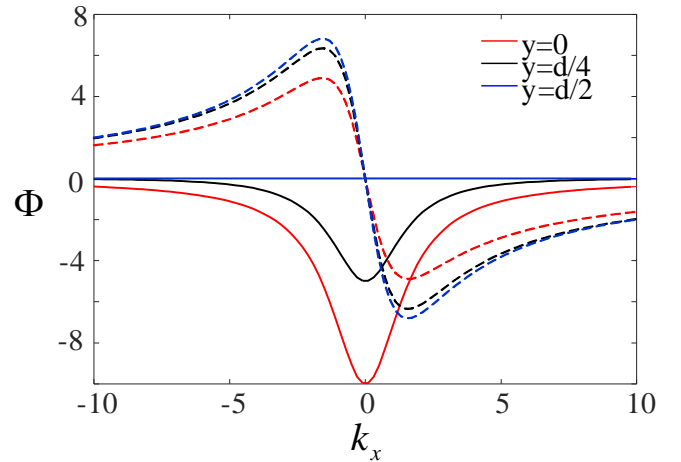

Fig. 6 $\text{Re}\{\Phi(k_x,y)\}$ (dashed line) and $\text{Im}\{\Phi(k_x,y)\}$ (solid line) for some values of $y$. The functions are calculated with (35) and (33) using (38) (the curves coincide for any $y$ and are not distinguishable).

### III. FIELDS AND IMPEDANCE

#### A. EM Fields

The fields on the transverse plane are obtained by differentiating the potential, namely, $\nabla_t\psi(x,y) = -\mathbf{e}_t(x,y)$,

where $\psi(x,y) = \varphi(x,y) + \mathrm{sgn}(x)V_0/2$ and $\varphi(x,y)$ is given by one of the alternative forms provided in section II. Being the mode TEM, the magnetic field is given by $\mathbf{h}_t(x,y) = \hat{\mathbf{z}} \times \mathbf{e}_t(x,y)/\zeta$ where $\zeta$ is the free space impedance. Correspondingly, the field spectra are found as $\mathbf{E}_t = (jk_x\hat{\mathbf{x}} - \hat{\mathbf{y}}\partial/\partial y)\Psi(k_x,y)$ and $\mathbf{H}_t = \hat{\mathbf{z}} \times \mathbf{E}_t/\zeta$ which can be rewritten using (33) and (38) as

$$\mathbf{E}_t(k_x,y) = -V_0\left[(\hat{\mathbf{x}} + j\hat{\mathbf{y}})\Psi^+ + (\hat{\mathbf{x}} - j\hat{\mathbf{y}})\Psi^-\right]$$
$$\mathbf{H}_t(x,y) = -\frac{1}{\zeta}V_0\left[(\hat{\mathbf{y}} - j\hat{\mathbf{x}})\Psi^+ + (\hat{\mathbf{y}} + j\hat{\mathbf{x}})\Psi^-\right] \quad (40)$$

where

$$\Psi^+(k_x,y) = \frac{\Gamma(3/4)}{\Gamma(1/4)} \frac{\Gamma\left(j\frac{k_x d}{2\pi} + \frac{1}{4}\right)}{\Gamma\left(j\frac{k_x d}{2\pi} + \frac{3}{4}\right)} \frac{\sinh(k_x y)}{\sinh(k_x d)} \quad (41)$$

$$\Psi^-(k_x,y) = \Psi^+(-k_x,(d-y)) \quad (42)$$

Equation (40) yields a decomposition of the fields in terms of left-handed (LH) and right-handed (RH) circularly polarized waves. The asymptotic behavior of the functions $\Psi^\pm(k_x,y)$ is of type $(k_x d)^{-1/2}$, which implies that the field is singular at the junctions between PEC and PMC boundary conditions with singularity of type $x^{-1/2}$.

### B. Characteristic impedance of the waveguide

Being the mode TEM, the wave impedance is coincident with the free space impedance $\zeta = 377\Omega$. Interestingly, the characteristic impedance $R = V_0/I_0$ of the waveguide is also equal to $\zeta$. To show this, we calculate the current $I_0$ by integrating the magnetic field on the PEC portion of the waveguide. The electric current density on $y=0$ is given by

$$I_0 = \int_{-\infty}^{0^-} \mathbf{j}(x) \cdot \mathbf{z}\,dx = -\frac{1}{\zeta}\int_{-\infty}^{0^-} \frac{\partial}{\partial x}\psi(x,0)dx$$
$$= -\frac{V_0}{\zeta}\sum_{n=0}^{\infty} \xi_n c_n \int_{-\infty}^{0^-} e^{-\xi_n|x|}dx = \frac{V_0}{\zeta}\sum_{n=0}^{\infty} c_n = \frac{V_0}{\zeta} \quad (43)$$

where the last equality is due to the property $\sum_{n=0}^{\infty} c_n = 1$. So, the characteristic impedance of the waveguide for this structure is equal to the wave impedance and to the free space impedance. This implies a power flow equal to $P = |V_0|^2/(2\zeta)$. Note that the characteristic impedance is independent of the distance between the plates $d$. This can be understood with a simple dimensional analysis noting that there is no other characteristic length scale in the problem, and thus the characteristic impedance must be totally independent of $d$.

## IV. PROTECTED PROPAGATION

Examples of protected propagation are shown in the following, with results obtained with CST Microwave Studio.

### A. Segmented PTD-waveguide

In this section, we study the propagation in a segmented PTD-waveguide with ideal PEC-PMC walls (Fig. 7). The distance $d$ between the two parallel walls has been set up as $d=1$cm, and the material in between the plates is air. This case serves to estimate the effect of repetitive 90°-bend discontinuities on the propagation. It is seen that the propagation is protected through the entire path with the reflection coefficient precisely zero at the input port over the entire bandwidth (in the CST simulation the reflection coefficient is on the order of -25dB due to numerical noise). Indeed, the segmented guide is PTD symmetric and for $\lambda > 4d$ the input and output ports support a single propagating mode (the TEM mode characterized in the previous sections). Hence, the system may be regarded as a two-port network and the fact that $s_{11} = s_{22} = 0$ guarantees that back-reflections are forbidden [16]. A snapshot of the vertical ($y$-component) of the electric field in the middle plane of the waveguide is presented in Fig. 7. The field is excited by a short vertical dipole radiating at 1GHz at the junction discontinuity of boundary conditions. The colored map in Fig. 7 shows a strong confinement of the field along the line discontinuity with no diffraction losses.

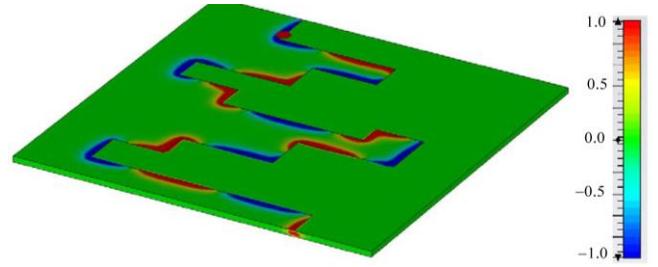

Fig. 7 Time snapshot of the $y$-component of the electric field excited by a point source in a segmented PTD-PPW waveguide. The represented area has electrical dimensions 2λ x 2λ.

### B. Open ended PEC-PMC waveguide

The effect of opening the PTD-PPW waveguide into free-space is investigated in Fig. 8. Being the free space a PTD-symmetric medium, we expect zero back-reflections at the input port (note that the input port supports a single propagating mode and the PTD invariance guarantees that $s_{11} = 0$).

Figure 8a shows the effect of a single PTD-PPW opened in free-space. The distance between the walls is $d=2$mm. A time snapshot of the magnitude of the electric field at 30GHz is visualized at the intermediate section between the two walls. The reflection coefficient has been found to be less than -20 dB all over the unimodal bandwidth. A second numerical example is obtained by pairing two close-by waveguide alternating PEC-PMC-PEC boundary conditions on one wall and PMC-PEC-PMC on the other wall. The distance between the two adjacent PMC-PEC junctions is 5mm. Figure 8b shows the distribution of the field magnitude when the two waveguides are excited by two modes in phase. Again, the PTD symmetry guarantees that the individual guides are matched (but they are not necessarily isolated, and hence now it is not guaranteed that all the incoming energy is radiated towards the free-space region). The numerically calculated active reflection coefficient (which includes the effect of numerical noise) is however below 16dB in this second setup, which indicates good port isolation. It is worth noting the good matching level that is obtained despite the small value of *d* in terms of the wavelength.

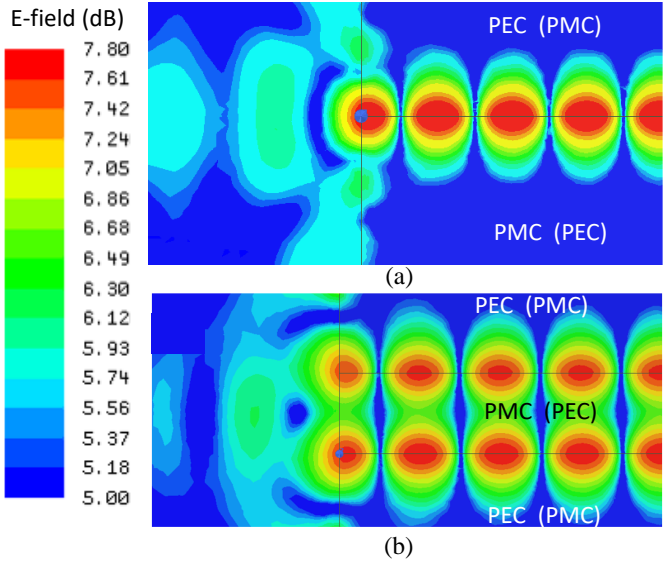

Fig. 8. a) PTD-PPW waveguide open in free space. (a) single PTD-PPW (b) double PTD-PPW fed by two TEM modes in phase. The black horizontal lines denote the positions of the junctions between the PEC and PMC boundaries; the vertical black line denotes the interface with the free space. The results are obtained using CST Microwave Studio.

## V. EXAMPLES OF PRACTICAL IMPLEMENTATION

A practical implementation of the PEC-PMC waveguide can be obtained by realizing equivalent boundary conditions through a bed of nails [20], [21]. A structure formed by a bed of nails top covered by a PEC wall (see inset "A" of Fig. 9a) exhibits an electromagnetic bandgap (EBG) [22]-[24]. Coupling this structure with an identical one with opposite position of the pins, generates a mode inside the bandgap. In the example of Fig. 9, the length of the nails is six times the distance *d* between the equivalent walls. The dispersion diagram is shown in Fig. 9a (calculated with CST Microwave Studio, case relevant to *d*=0.5mm). This diagram was obtained by solving the eigenvalue problem for a section of the structure with lateral pec boundary conditions at a sufficiently large distance *D* from *x*=0. In the simulation, we chose *D* such that the dominant mode (*n*=0 in (30)) is attenuated by a factor 10000 (this leads to *D*=5.86*d*). With the selected value of *d*=0.5mm, the bandgap of each half of the waveguide occurs in the spectral range [25-40] GHz (grey region in fig. 9a). The two boundaries of the bandgap correspond to 6*d*=[0.25-0.4]λ or equivalently *d*=[0.04-0.0667]λ. In the bandgap of the top covered bed of nail ("A" in inset), a quasi-TEM mode is found for the composite structure emulating the PTD waveguide ("B" in the inset). This mode is quite non-dispersive in the region 30-40 GHz. The dispersion curves below 25GHz are associated with spurious modes that travel inside the multi-connected wire domain constituted by the nails-region. They are present for both structures A and B in the simulation.

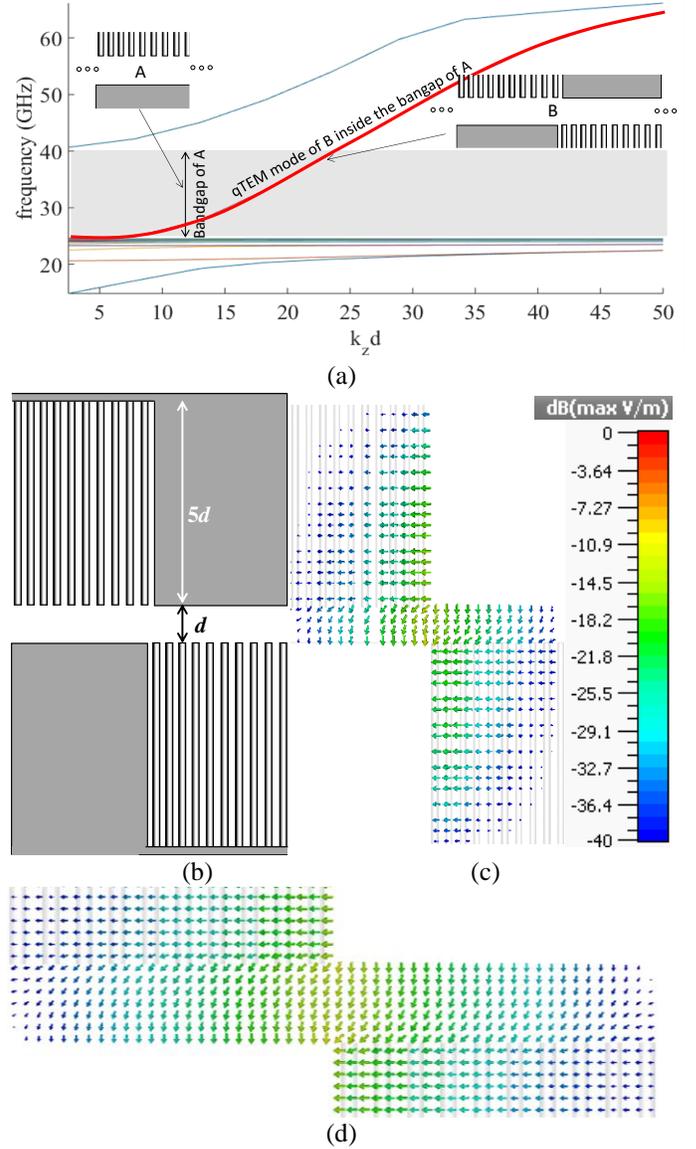

Fig. 9. (a) Dispersion diagram of modes in the PTD-waveguide with the PMC walls implemented using a bed of nails. (b) Cross-section of the PTD waveguide (c) Distribution of the electric field associated with the quasi-TEM mode (d) Zoom of the central part.

The transverse field associated with the dominant quasi-TEM mode (Fig. 9c and zoom in Fig. 9d) closely resembles the one associated with the ideal PTD waveguide (see Fig. 1b).

A more compact structure may be obtained by using the mushroom ground plane [25]. An example is illustrated in Fig. 10. The mushroom structure (see Fig. 10a) is realized with a dielectric slab of thickness 0.5mm and relative permittivity equal to 6, and the distance between the equivalent walls is 0.5mm. Fig. 10b shows the dispersion diagram of the first 14 modes supported by the structure, calculated with CST Microwave Studio[TM]. As it can be seen, there is a unimodal region between 28.2GHz and 41.4GHz, which falls in the EBG of each half of the waveguide. The mode supported in this region is quasi-TEM, and its electric field distribution at 35GHz is shown in Fig. 10c. Also in this case, the field distribution is somewhat similar to the one in the ideal PTD waveguide (see the inset of Fig. 10c).

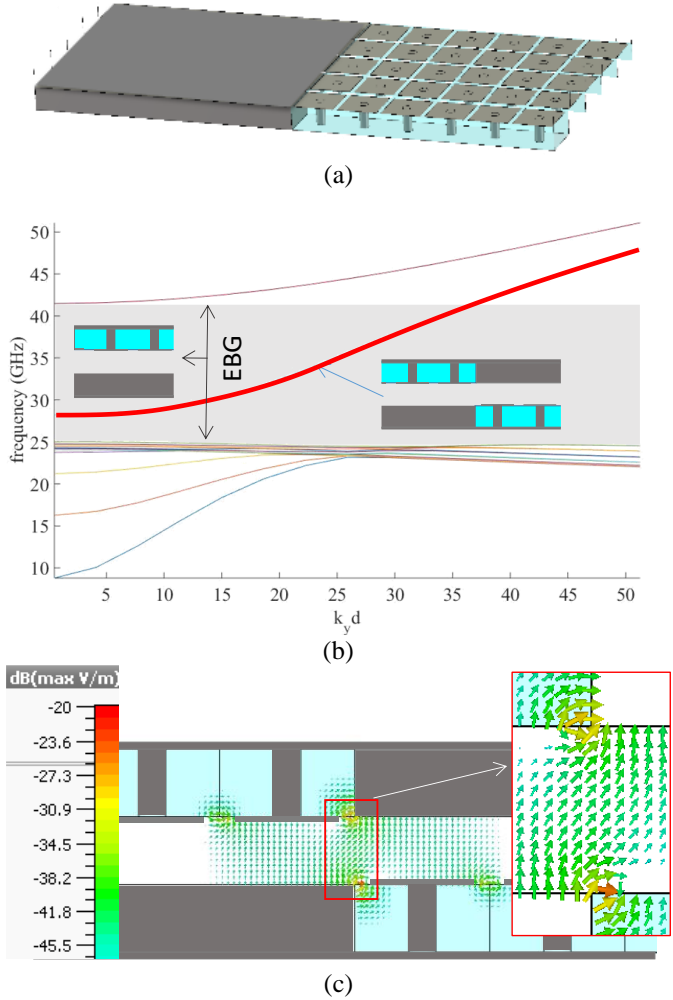

Fig. 10. PTD-waveguide with the PMC walls implemented using a mushroom ground plane. (a) Geometry of the lower half of the structure (b) Dispersion diagram (c) Distribution of the electric field associated with the quasi-TEM mode.

## VI. CONCLUSION

The exact solution of a PTD symmetric structure constituted by a Perfectly Electric-Perfectly Magnetic (PEC-PMC) parallel plate waveguide (PPW) has been found. This waveguide supports unimodal TEM edge mode propagation protected against back-scattering deformations and defects that maintain a PTD symmetry. The exact TEM solution is found in analytical form by using three different methods, namely by conformal mapping, by mode-matching, and by Fourier-transform method. It is found that the three solutions are coincident and each of them highlights different aspects of the structure of the modal fields. The characteristic impedance of the TEM mode coincides with the wave impedance of the mode and with the one of the free space. It is also shown through numerical simulations that the mode propagation is robust with respect to deformations such as 90° bends and discontinuity in free-space, confirming that PTD symmetric reciprocal structures are matched at all ports. Implementation of the PMC boundaries via both a bed of nails and a mushroom structure is also proposed, showing a quite linear dispersion characteristic.

## APPENDIX

### A. Derivation of the linear system for the coefficients $a_n$

The linear systems $\sum_{n=0}^{\infty}\zeta_{mn}^{(i)}a_n = V_m^{(i)}$ $m=0, 1, 2....$, for $i=0,1,2,...$ in (22) are obtained by imposing the continuity of the potential $\psi$ and of its derivatives at $x=0$ through testing with the functions $\sin(\alpha_m y)$.

The coefficients of the linear systems are calculated as

$$V_m^{(0)} = V_0 \int_0^d \sin(\alpha_m y) dy = V_0 \frac{d}{(m\pi + \pi/2)} \qquad (44)$$

$$V_m^{(i)} = 0, \; i=1,2 \qquad (45)$$

$$\zeta_{mn}^{(0)} = \int_0^d \left[\sin(\alpha_n y) + (-1)^n \cos(\alpha_n y)\right]\sin(\alpha_m y) dy$$
$$= \frac{d}{2}\delta_{nm} + \frac{d}{2}\left[\frac{(-1)^m(2n+1)-(-1)^n(2m+1)}{\pi(n+1+m)(n-m)}\right] \qquad (46)$$

$$\zeta_{mn}^{(1)} = \int_0^d \alpha_n \left[\cos(\alpha_n y) - (-1)^n \sin(\alpha_n y)\right]\sin(\alpha_m y) dy$$
$$= -\frac{d}{2}\alpha_m \delta_{nm} + \frac{d}{2}\alpha_n \left[\frac{(-1)^m(2n+1)-(-1)^n(2m+1)}{\pi(n+1+m)(n-m)}\right] \qquad (47)$$

$$\zeta_{mn}^{(2)} = \int_0^d \alpha_n \left[\sin(\alpha_n y) - (-1)^n \cos(\alpha_n y)\right]\sin(\alpha_m y) dy$$
$$= \alpha_m \frac{d}{2}\delta_{nm} - \frac{d}{2}\alpha_n \left[\frac{(-1)^n(2m+1)-(-1)^m(2n+1)}{\pi(n+1+m)(m-n)}\right] \qquad (48)$$

### B. Determination of the function f(z)

The function $f(z)$ in (26) should have the form

$$f(z) = p(z) \frac{\prod_{m=1}^{\infty}\left(1-\frac{z}{2m-1}\right)e^{z/2m}}{z\prod_{n=1}^{\infty}\left(1-\frac{z}{2n}\right)e^{z/2n}} \qquad (49)$$

where the exponential terms have been introduced in order to ensure convergence of the infinite product. Comparing it with the definition of the gamma function, one can find that ([27], p. 59)

$$\prod_{n=1}^{\infty}\left(1-\frac{z}{2n}\right)e^{z/2n} = -\frac{2e^{\gamma z/2}}{\Gamma\left(-\frac{z}{2}\right)z}$$

$$\prod_{m=1}^{\infty}\left(1-\frac{z}{2m-1}\right)e^{z/2m} = \frac{e^{\gamma z/2}\sqrt{\pi}}{\Gamma\left(-\frac{z}{2}+\frac{1}{2}\right)} \tag{50}$$

from which one has $p(z) = const.$ therefore obtaining (26). In order to find the residues, one can use the mathematical identities

$$\Gamma\left(-\frac{z}{2}\right)\Gamma\left(\frac{z}{2}\right) = -\frac{2\pi}{z\sin\left(\frac{z\pi}{2}\right)}$$

$$\Gamma\left(-\frac{z}{2}+\frac{1}{2}\right)\Gamma\left(\frac{z}{2}+\frac{1}{2}\right) = \frac{\pi}{\cos\left(\frac{z\pi}{2}\right)} \tag{51}$$

from which

$$f(z) = C\frac{\Gamma\left(-\frac{z}{2}\right)}{\Gamma\left(-\frac{z}{2}+\frac{1}{2}\right)} = -C\frac{\Gamma\left(\frac{z}{2}+\frac{1}{2}\right)}{z\Gamma\left(\frac{z}{2}\right)}2\cot\left(\frac{z\pi}{2}\right) \tag{52}$$

Note that the above can be also rewritten by using the identity

$$\Gamma\left(\frac{z}{2}+\frac{1}{2}\right)\Gamma\left(\frac{z}{2}\right) = 2\sqrt{\pi}e^{-z\ln 2}\Gamma(z) \tag{53}$$

as

$$f(z) = -C\frac{2\sqrt{\pi}e^{-z\ln 2}\Gamma(z)}{z\left(\Gamma\left(\frac{z}{2}\right)\right)^2}2\cot\left(\frac{z\pi}{2}\right) \tag{54}$$

which is the function introduced by Collin for solving a boundary value problem of step-discontinuity [19]. The residue of the function $f(z)$ at the pole $z = 2n$ can be found as

$$\lim_{z \to 2n} f(z)(z-2n) = R_n = K'\frac{\Gamma(n+1/2)}{n\Gamma(n)} \quad n = 1, 2, ...$$

$$\lim_{z \to 0} f(z)z = R_0 = K'\sqrt{\pi} \quad n = 0 \tag{55}$$

where we have used $\lim_{z\to 0}[z\Gamma(z)] = 1$, $\Gamma(1/2) = \sqrt{\pi}$ and $K' = 2C/\pi$. From here, one can find the coefficients $a_{2n}$. Observing that $\Gamma(n) = (n-1)!$ and, from (44), that $\Gamma(n+1/2) = \sqrt{\pi}(2n)!/(4^n n!)$ one has

$$R_n = K\frac{\sqrt{\pi}(2n)!}{(n!)^2 4^n} \quad n = 1, 2, 3, ... \tag{56}$$

## C. Construction of the function A(s)

To construct the function $A(s)$ we use the relation $A(s) = N(s)/D(s)$ with

$$N(s) = \prod_{m=1}^{\infty}\left(1-\frac{s}{j(2\pi m - \pi/2)}\right)e^{s/(j2\pi m)} = \frac{2e^{\gamma s/(j2\pi)}\Gamma\left(\frac{3}{4}\right)}{\Gamma\left(\frac{js}{2\pi}+\frac{3}{4}\right)}$$

$$D(s) = \prod_{n=1}^{\infty}\left(1-\frac{s}{j(2\pi n - 3\pi/2)}\right)e^{s/(j2\pi n)} = \frac{2e^{\gamma s/(j2\pi)}\Gamma\left(\frac{1}{4}\right)}{\Gamma\left(\frac{js}{2\pi}+\frac{1}{4}\right)} \tag{57}$$

which can be found as the general form of the expression reported in [26] and leads to (38).

## ACKNOWLEDGMENTS

M.S. was partially supported by the European Regional Development Fund (FEDER), through the Competitiveness and Internationalization Operational Programme (COMPETE 2020) of the Portugal 2020 framework, Project, RETIOT, POCI-01-0145-FEDER-016432, and by Fundação para Ciência e a Tecnologia (FCT) under projects PTDC/EEITEL/ 4543/2014 and UID/EEA/50008/2017. The authors thank the collaboration Iuliia Rybalko in obtaining the data for Fig. 7.